\documentclass[letterpaper,compsoc,twoside]{IEEEtran}
\usepackage{fixltx2e} % LaTeX patches, \textsubscript
\usepackage{cmap} % fix search and cut-and-paste in Acrobat
\usepackage{ifthen}
\usepackage[T1]{fontenc}
\usepackage[utf8]{inputenc}
\usepackage{amsmath}

\usepackage[font={small,it},labelfont=bf]{caption}
\usepackage{float}

\usepackage{scipy}
\makeatletter
\def\PY@reset{\let\PY@it=\relax \let\PY@bf=\relax%
    \let\PY@ul=\relax \let\PY@tc=\relax%
    \let\PY@bc=\relax \let\PY@ff=\relax}
\def\PY@tok#1{\csname PY@tok@#1\endcsname}
\def\PY@toks#1+{\ifx\relax#1\empty\else%
    \PY@tok{#1}\expandafter\PY@toks\fi}
\def\PY@do#1{\PY@bc{\PY@tc{\PY@ul{%
    \PY@it{\PY@bf{\PY@ff{#1}}}}}}}
\def\PY#1#2{\PY@reset\PY@toks#1+\relax+\PY@do{#2}}

\expandafter\def\csname PY@tok@nc\endcsname{\let\PY@bf=\textbf\def\PY@tc##1{\textcolor[rgb]{0.05,0.52,0.71}{##1}}}
\expandafter\def\csname PY@tok@vg\endcsname{\def\PY@tc##1{\textcolor[rgb]{0.73,0.38,0.84}{##1}}}
\expandafter\def\csname PY@tok@vc\endcsname{\def\PY@tc##1{\textcolor[rgb]{0.73,0.38,0.84}{##1}}}
\expandafter\def\csname PY@tok@cs\endcsname{\def\PY@tc##1{\textcolor[rgb]{0.25,0.50,0.56}{##1}}\def\PY@bc##1{\setlength{\fboxsep}{0pt}\colorbox[rgb]{1.00,0.94,0.94}{\strut ##1}}}
\expandafter\def\csname PY@tok@ni\endcsname{\let\PY@bf=\textbf\def\PY@tc##1{\textcolor[rgb]{0.84,0.33,0.22}{##1}}}
\expandafter\def\csname PY@tok@kn\endcsname{\let\PY@bf=\textbf\def\PY@tc##1{\textcolor[rgb]{0.00,0.44,0.13}{##1}}}
\expandafter\def\csname PY@tok@ne\endcsname{\def\PY@tc##1{\textcolor[rgb]{0.00,0.44,0.13}{##1}}}
\expandafter\def\csname PY@tok@nl\endcsname{\let\PY@bf=\textbf\def\PY@tc##1{\textcolor[rgb]{0.00,0.13,0.44}{##1}}}
\expandafter\def\csname PY@tok@sd\endcsname{\let\PY@it=\textit\def\PY@tc##1{\textcolor[rgb]{0.25,0.44,0.63}{##1}}}
\expandafter\def\csname PY@tok@ss\endcsname{\def\PY@tc##1{\textcolor[rgb]{0.32,0.47,0.09}{##1}}}
\expandafter\def\csname PY@tok@s2\endcsname{\def\PY@tc##1{\textcolor[rgb]{0.25,0.44,0.63}{##1}}}
\expandafter\def\csname PY@tok@c\endcsname{\let\PY@it=\textit\def\PY@tc##1{\textcolor[rgb]{0.25,0.50,0.56}{##1}}}
\expandafter\def\csname PY@tok@nd\endcsname{\let\PY@bf=\textbf\def\PY@tc##1{\textcolor[rgb]{0.33,0.33,0.33}{##1}}}
\expandafter\def\csname PY@tok@gh\endcsname{\let\PY@bf=\textbf\def\PY@tc##1{\textcolor[rgb]{0.00,0.00,0.50}{##1}}}
\expandafter\def\csname PY@tok@m\endcsname{\def\PY@tc##1{\textcolor[rgb]{0.13,0.50,0.31}{##1}}}
\expandafter\def\csname PY@tok@kc\endcsname{\let\PY@bf=\textbf\def\PY@tc##1{\textcolor[rgb]{0.00,0.44,0.13}{##1}}}
\expandafter\def\csname PY@tok@o\endcsname{\def\PY@tc##1{\textcolor[rgb]{0.40,0.40,0.40}{##1}}}
\expandafter\def\csname PY@tok@c1\endcsname{\let\PY@it=\textit\def\PY@tc##1{\textcolor[rgb]{0.25,0.50,0.56}{##1}}}
\expandafter\def\csname PY@tok@sb\endcsname{\def\PY@tc##1{\textcolor[rgb]{0.25,0.44,0.63}{##1}}}
\expandafter\def\csname PY@tok@il\endcsname{\def\PY@tc##1{\textcolor[rgb]{0.13,0.50,0.31}{##1}}}
\expandafter\def\csname PY@tok@mo\endcsname{\def\PY@tc##1{\textcolor[rgb]{0.13,0.50,0.31}{##1}}}
\expandafter\def\csname PY@tok@se\endcsname{\let\PY@bf=\textbf\def\PY@tc##1{\textcolor[rgb]{0.25,0.44,0.63}{##1}}}
\expandafter\def\csname PY@tok@si\endcsname{\let\PY@it=\textit\def\PY@tc##1{\textcolor[rgb]{0.44,0.63,0.82}{##1}}}
\expandafter\def\csname PY@tok@nb\endcsname{\def\PY@tc##1{\textcolor[rgb]{0.00,0.44,0.13}{##1}}}
\expandafter\def\csname PY@tok@ch\endcsname{\let\PY@it=\textit\def\PY@tc##1{\textcolor[rgb]{0.25,0.50,0.56}{##1}}}
\expandafter\def\csname PY@tok@sc\endcsname{\def\PY@tc##1{\textcolor[rgb]{0.25,0.44,0.63}{##1}}}
\expandafter\def\csname PY@tok@mh\endcsname{\def\PY@tc##1{\textcolor[rgb]{0.13,0.50,0.31}{##1}}}
\expandafter\def\csname PY@tok@gi\endcsname{\def\PY@tc##1{\textcolor[rgb]{0.00,0.63,0.00}{##1}}}
\expandafter\def\csname PY@tok@k\endcsname{\let\PY@bf=\textbf\def\PY@tc##1{\textcolor[rgb]{0.00,0.44,0.13}{##1}}}
\expandafter\def\csname PY@tok@kr\endcsname{\let\PY@bf=\textbf\def\PY@tc##1{\textcolor[rgb]{0.00,0.44,0.13}{##1}}}
\expandafter\def\csname PY@tok@nf\endcsname{\def\PY@tc##1{\textcolor[rgb]{0.02,0.16,0.49}{##1}}}
\expandafter\def\csname PY@tok@err\endcsname{\def\PY@bc##1{\setlength{\fboxsep}{0pt}\fcolorbox[rgb]{1.00,0.00,0.00}{1,1,1}{\strut ##1}}}
\expandafter\def\csname PY@tok@s1\endcsname{\def\PY@tc##1{\textcolor[rgb]{0.25,0.44,0.63}{##1}}}
\expandafter\def\csname PY@tok@mi\endcsname{\def\PY@tc##1{\textcolor[rgb]{0.13,0.50,0.31}{##1}}}
\expandafter\def\csname PY@tok@nt\endcsname{\let\PY@bf=\textbf\def\PY@tc##1{\textcolor[rgb]{0.02,0.16,0.45}{##1}}}
\expandafter\def\csname PY@tok@sh\endcsname{\def\PY@tc##1{\textcolor[rgb]{0.25,0.44,0.63}{##1}}}
\expandafter\def\csname PY@tok@cm\endcsname{\let\PY@it=\textit\def\PY@tc##1{\textcolor[rgb]{0.25,0.50,0.56}{##1}}}
\expandafter\def\csname PY@tok@gu\endcsname{\let\PY@bf=\textbf\def\PY@tc##1{\textcolor[rgb]{0.50,0.00,0.50}{##1}}}
\expandafter\def\csname PY@tok@mb\endcsname{\def\PY@tc##1{\textcolor[rgb]{0.13,0.50,0.31}{##1}}}
\expandafter\def\csname PY@tok@gp\endcsname{\let\PY@bf=\textbf\def\PY@tc##1{\textcolor[rgb]{0.78,0.36,0.04}{##1}}}
\expandafter\def\csname PY@tok@vi\endcsname{\def\PY@tc##1{\textcolor[rgb]{0.73,0.38,0.84}{##1}}}
\expandafter\def\csname PY@tok@kp\endcsname{\def\PY@tc##1{\textcolor[rgb]{0.00,0.44,0.13}{##1}}}
\expandafter\def\csname PY@tok@s\endcsname{\def\PY@tc##1{\textcolor[rgb]{0.25,0.44,0.63}{##1}}}
\expandafter\def\csname PY@tok@cpf\endcsname{\let\PY@it=\textit\def\PY@tc##1{\textcolor[rgb]{0.25,0.50,0.56}{##1}}}
\expandafter\def\csname PY@tok@cp\endcsname{\def\PY@tc##1{\textcolor[rgb]{0.00,0.44,0.13}{##1}}}
\expandafter\def\csname PY@tok@kd\endcsname{\let\PY@bf=\textbf\def\PY@tc##1{\textcolor[rgb]{0.00,0.44,0.13}{##1}}}
\expandafter\def\csname PY@tok@kt\endcsname{\def\PY@tc##1{\textcolor[rgb]{0.56,0.13,0.00}{##1}}}
\expandafter\def\csname PY@tok@ow\endcsname{\let\PY@bf=\textbf\def\PY@tc##1{\textcolor[rgb]{0.00,0.44,0.13}{##1}}}
\expandafter\def\csname PY@tok@nn\endcsname{\let\PY@bf=\textbf\def\PY@tc##1{\textcolor[rgb]{0.05,0.52,0.71}{##1}}}
\expandafter\def\csname PY@tok@gd\endcsname{\def\PY@tc##1{\textcolor[rgb]{0.63,0.00,0.00}{##1}}}
\expandafter\def\csname PY@tok@na\endcsname{\def\PY@tc##1{\textcolor[rgb]{0.25,0.44,0.63}{##1}}}
\expandafter\def\csname PY@tok@mf\endcsname{\def\PY@tc##1{\textcolor[rgb]{0.13,0.50,0.31}{##1}}}
\expandafter\def\csname PY@tok@ge\endcsname{\let\PY@it=\textit}
\expandafter\def\csname PY@tok@gr\endcsname{\def\PY@tc##1{\textcolor[rgb]{1.00,0.00,0.00}{##1}}}
\expandafter\def\csname PY@tok@w\endcsname{\def\PY@tc##1{\textcolor[rgb]{0.73,0.73,0.73}{##1}}}
\expandafter\def\csname PY@tok@nv\endcsname{\def\PY@tc##1{\textcolor[rgb]{0.73,0.38,0.84}{##1}}}
\expandafter\def\csname PY@tok@sr\endcsname{\def\PY@tc##1{\textcolor[rgb]{0.14,0.33,0.53}{##1}}}
\expandafter\def\csname PY@tok@gt\endcsname{\def\PY@tc##1{\textcolor[rgb]{0.00,0.27,0.87}{##1}}}
\expandafter\def\csname PY@tok@sx\endcsname{\def\PY@tc##1{\textcolor[rgb]{0.78,0.36,0.04}{##1}}}
\expandafter\def\csname PY@tok@no\endcsname{\def\PY@tc##1{\textcolor[rgb]{0.38,0.68,0.84}{##1}}}
\expandafter\def\csname PY@tok@gs\endcsname{\let\PY@bf=\textbf}
\expandafter\def\csname PY@tok@go\endcsname{\def\PY@tc##1{\textcolor[rgb]{0.20,0.20,0.20}{##1}}}
\expandafter\def\csname PY@tok@bp\endcsname{\def\PY@tc##1{\textcolor[rgb]{0.00,0.44,0.13}{##1}}}

\def\PYZus{\char`\_}
\def\PYZob{\char`\{}
\def\PYZcb{\char`\}}

\def\PYZlt{\char`\<}
\def\PYZgt{\char`\>}

\def\PYZpc{\char`\%}

\def\PYZhy{\char`\-}
\def\PYZsq{\char`\'}
\def\PYZdq{\char`\"}

\makeatother

\providecommand*{\DUfootnotemark}[3]{%
  \raisebox{1em}{\hypertarget{#1}{}}%
  \hyperlink{#2}{\textsuperscript{#3}}%
}
\providecommand{\DUfootnotetext}[4]{%
  \begingroup%
  \renewcommand{\thefootnote}{%
    \protect\raisebox{1em}{\protect\hypertarget{#1}{}}%
    \protect\hyperlink{#2}{#3}}%
  \footnotetext{#4}%
  \endgroup%
}

\providecommand*{\DUrole}[2]{%
  \ifcsname DUrole#1\endcsname%
    \csname DUrole#1\endcsname{#2}%
  \else% backwards compatibility: try \docutilsrole#1{#2}
    \ifcsname docutilsrole#1\endcsname%
      \csname docutilsrole#1\endcsname{#2}%
    \else%
      #2%
    \fi%
  \fi%
}

\providecommand*{\DUroletitlereference}[1]{\textsl{#1}}

\ifthenelse{\isundefined{\hypersetup}}{
  \usepackage[colorlinks=true,linkcolor=blue,urlcolor=blue]{hyperref}
  \urlstyle{same} % normal text font (alternatives: tt, rm, sf)
}{}

\begin{document}
\newcounter{footnotecounter}\title{cesium: Open-Source Platform for Time-Series Inference}\author{Brett Naul$^{\setcounter{footnotecounter}{3}\fnsymbol{footnotecounter}\setcounter{footnotecounter}{1}\fnsymbol{footnotecounter}}$%
          \setcounter{footnotecounter}{1}\thanks{\fnsymbol{footnotecounter} %
          Corresponding author: \protect\href{mailto:bnaul@berkeley.edu}{bnaul@berkeley.edu}}\setcounter{footnotecounter}{3}\thanks{\fnsymbol{footnotecounter} University of California, Berkeley}, Stéfan van der Walt$^{\setcounter{footnotecounter}{3}\fnsymbol{footnotecounter}}$, Arien Crellin-Quick$^{\setcounter{footnotecounter}{3}\fnsymbol{footnotecounter}}$, Joshua S. Bloom$^{\setcounter{footnotecounter}{4}\fnsymbol{footnotecounter}\setcounter{footnotecounter}{3}\fnsymbol{footnotecounter}}$\setcounter{footnotecounter}{4}\thanks{\fnsymbol{footnotecounter} Lawrence Berkeley National Laboratory}, Fernando Pérez$^{\setcounter{footnotecounter}{4}\fnsymbol{footnotecounter}\setcounter{footnotecounter}{3}\fnsymbol{footnotecounter}}$\thanks{%

          \noindent%
          Copyright\,\copyright\,2016 Brett Naul et al. This is an open-access article distributed under the terms of the Creative Commons Attribution License, which permits unrestricted use, distribution, and reproduction in any medium, provided the original author and source are credited.%
        }}\maketitle
          \renewcommand{\leftmark}{PROC. OF THE 15th PYTHON IN SCIENCE CONF. (SCIPY 2016)}
          \renewcommand{\rightmark}{CESIUM: OPEN-SOURCE PLATFORM FOR TIME-SERIES INFERENCE}

\InputIfFileExists{page_numbers.tex}{}{}
\newcommand*{\docutilsroleref}{\ref}
\newcommand*{\docutilsrolelabel}{\label}
\providecommand*\DUrolecite[1]{\cite{#1}}
\begin{abstract}Inference on time series data is a common requirement in many scientific
disciplines and internet of things (IoT) applications, yet there are few
resources available to domain scientists to easily, robustly, and repeatably
build such complex inference workflows: traditional statistical
models of time series are often too rigid to explain complex time domain
behavior, while popular machine learning packages require already-featurized
dataset inputs. Moreover, the software engineering tasks required to
instantiate the computational platform are daunting. \texttt{cesium} is an
end-to-end time series analysis framework, consisting of a Python library as
well as a web front-end interface, that allows researchers to featurize raw
data and apply modern machine learning techniques in a simple, reproducible,
and extensible way. Users can apply out-of-the-box feature engineering
workflows as well as save and replay their own analyses. Any steps taken in
the front end can also be exported to a Jupyter notebook, so users can
iterate between possible models within the front end and then fine-tune their
analysis using the additional capabilities of the back-end library. The
open-source packages make us of many use modern Python toolkits, including
\texttt{xarray}, \texttt{dask}, Celery, Flask, and \texttt{scikit-learn}.\end{abstract}

\subsection{Introduction%
  \label{introduction}%
}

From the reading of electroencephalograms (EEGs) to earthquake seismograms to
light curves of astronomical variable stars, gleaning insight from time series
data has been central to a broad range of scientific disciplines.
When simple analytical thresholds or models suffice, technicians and experts can
be easily removed from the process of inspection and discovery by employing
custom algorithms. But when dynamical systems are not easily modeled (e.g.,
through physics-based models or standard regression techniques), classification
and anomaly detection have traditionally been reserved for the domain expert:
digitally recorded data are visually scanned to ascertain the nature of the time
variability and find important (perhaps life-threatening) outliers. \emph{Does this
person have an irregular heartbeat? What type of supernova occurred in that
galaxy?} Even in the presence of sensor noise and intrinsic diversity of the
samples, well-trained domain specialists show a remarkable ability to make
discerning statements about complex data.

In an era when more time series data are being collected than can be visually
inspected by domain experts, computational frameworks must necessarily
act as human surrogates. Capturing the subtleties that domain experts intuit in
time series data (let alone besting the experts) is a non-trivial task.
In this respect, machine learning has already been used to great success in
several disciplines, including text classification, image retrieval,
segmentation of remote sensing data, internet traffic classification, video
analysis, and classification of medical data. Even if the results are similar,
some obvious advantages over human involvement are that machine learning
algorithms are tunable, repeatable, and deterministic. A computational framework
built with elasticity can scale, whereas experts (and even crowdsourcing)
cannot.

Despite the importance of time series in scientific research, there are few
resources available that allow domain scientists to easily build robust
computational inference workflows for their own time series data, let alone
data gathered more broadly in their field. The difficulties involved in
constructing such a framework can often greatly outweigh those of analyzing the
data itself:\begin{quotation}%
\begin{quote}

It may be surprising to the academic community to know that only a tiny
fraction of the code in many machine learning systems is actually doing
\textquotedbl{}machine learning\textquotedbl{}...a mature system might end up
being (at most) 5\% machine learning code and (at least) 95\% glue code.
\DUrole{cite}{scu2014}
\end{quote}
\end{quotation}

Even if a domain scientist works closely with machine learning experts, the
software engineering requirements can be daunting. It is our opinion that being
a modern data-driven scientist should not require an army of software engineers,
machine learning experts, statisticians and production operators. \texttt{cesium}
\DUrole{cite}{cesium} was created to allow domain experts to focus on the inference
questions at hand rather than needing to assemble a complete engineering
project.

The analysis workflow of \texttt{cesium} can be used in two forms: a web front end
which allows researchers to upload their data, perform analyses, and visualize
their models all within the browser; and a Python library which exposes more
flexible interfaces to the same analysis tools. The web front end is designed to
handle many of the more cumbersome aspects of machine learning analysis,
including data uploading and management, scaling of computational resources, and
tracking of results from previous experiments. The Python library is used within
the web back end for the main steps of the analysis workflow: extracting
features from raw time series, building models from these features, and
generating predictions. The library also supplies data structures for storing
time series (including support for irregularly-sampled time series and
measurement errors), features, and other relevant metadata.

In the next section, we describe a few motivating examples of scientific time
series analysis problems. The subsequent sections describe in detail the
\texttt{cesium} library and web front end, including the different pieces of
functionality provided and various design questions and decisions that arose
during the development process. Finally, we present an end-to-end analysis of an
EEG seizure dataset, first using the Python library and then via the web front
end.

\subsection{Example time series machine learning problems%
  \label{example-time-series-machine-learning-problems}%
}

\texttt{cesium} was designed with several time series inference use cases across various
scientific disciplines in mind.\newcounter{listcnt0}
\begin{list}{\arabic{listcnt0}.}
{
\usecounter{listcnt0}
\setlength{\rightmargin}{\leftmargin}
}

\item 

\textbf{Astronomical time series classification.} Beginning in 2020, the Large
Synoptic Survey Telescope (LSST) will survey the entire night’s sky every few
days producing high-quality time series data on approximately 800 million
transient events and sources with variable brightness (Figure \DUrole{ref}{astro}
depicts the brightness of several types of star over the course of several
years) \DUrole{cite}{lsst2009}. Much of the best science in the time domain (e.g.,
the discovery of the accelerating universe and dark energy using Type Ia
supernovae \DUrole{cite}{perlmutter1999,riess1998}) consists of first identifying
possible phenomena of interest using broad data mining approaches and
following up by collecting more detailed data using other, more precise
observational tools. For many transient events, the time scale during which
observations can be collected can be on the order of days or hours. Not
knowing which of the millions of variable sources to examine more closely
with larger telescopes and specialized instruments is tantamount to not
having discovered those sources at all. Discoveries must be identified
quickly or in real time so that informed decisions can be made about how best
to allocate additional observational resources.\end{list}
\begin{figure}[]\noindent\makebox[\columnwidth][c]{\includegraphics[width=\columnwidth]{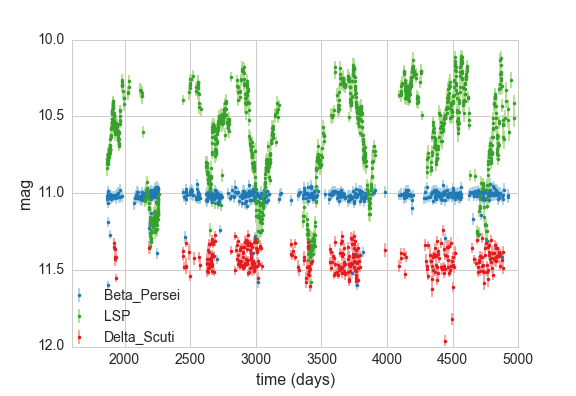}}
\caption{Typical data for a classification task on variable stars from the All Sky
Automated Survey; shown are flux measurements for three stars
irregularly sampled in time \DUrole{cite}{richards2012}. \DUrole{label}{astro}}
\end{figure}\setcounter{listcnt0}{0}
\begin{list}{\arabic{listcnt0}.}
{
\usecounter{listcnt0}
\addtocounter{listcnt0}{1}
\setlength{\rightmargin}{\leftmargin}
}

\item 

\textbf{Neuroscience time series classification.}
that might need to be classified in order to make treatment decisions.
Neuroscience experiments now produce vast amounts of time series data that
can have entirely different structures and spatial/temporal resolutions,
depending on the recording technique.
Figure \DUrole{ref}{eeg} shows an example of different types of EEG signals
The neuroscience community is turning to the use of large-scale machine
learning tools to extract insight from large, complex datasets
\DUrole{cite}{lotte2007}. However, the community lacks tools to validate and compare
data analysis approaches in a robust, efficient and reproducible manner: even
recent expert reviews on the matter leave many of these critical
methodological questions open for the user to explore in an ad hoc way and
with little principled guidance \DUrole{cite}{perez2007}.\end{list}
\begin{figure}[]\noindent\makebox[\columnwidth][c]{\includegraphics[width=\columnwidth]{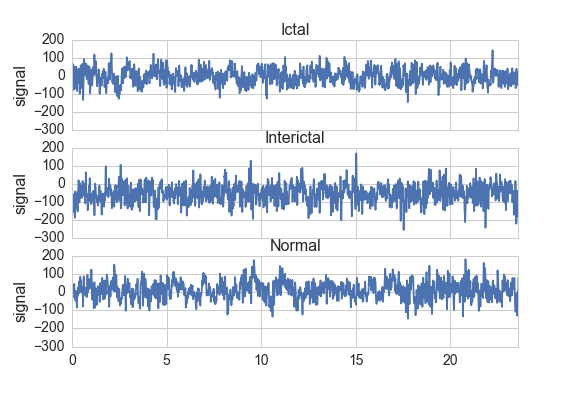}}
\caption{EEG signals from patients with epilepsy \DUrole{cite}{andrzejak2001}. \DUrole{label}{eeg}}
\end{figure}\setcounter{listcnt0}{0}
\begin{list}{\arabic{listcnt0}.}
{
\usecounter{listcnt0}
\addtocounter{listcnt0}{2}
\setlength{\rightmargin}{\leftmargin}
}

\item 

\textbf{Earthquake detection, characterization and warning.} Earthquake early
warning (EEW) systems are currently in operation in Japan, Mexico, Turkey,
Taiwan and Romania \DUrole{cite}{allen2009} and are under development in the US
\DUrole{cite}{brown2011}. These systems have employed sophisticated remote sensors,
real-time connectivity to major broadcast outlets (such as TV and radio), and
have a growing resumé of successful rapid assessment of threat levels to
populations and industry. Traditionally these warning systems trigger from
data obtained by high-quality seismic networks with sensors placed every \textasciitilde{}10
km. Today, however, accelerometers are embedded in many consumer electronics
including computers and smartphones. There is tremendous potential to improve
earthquake detection methods using streaming classification analysis both
using traditional network data and also harnessing massive data from consumer
electronics.\end{list}

\subsection{Simple and reproducible workflows%
  \label{simple-and-reproducible-workflows}%
}

In recent years, there has been rapid growth in the availability of open-source
tools that implement a wide variety of machine learning algorithms: packages
within the R \DUrole{cite}{team2013} and Python programming languages
\DUrole{cite}{pedregosa2011}, standalone Java-based packages such as Moa
\DUrole{cite}{bifet2010} and Weka \DUrole{cite}{hall2009}, and online webservices such as the
Google Prediction API, to name a few. To a domain scientist that does not have
formal training in machine learning, however, the availability of such packages
is both a blessing and a curse. On one hand, most machine learning algorithms
are now widely accessible to all researchers. At the same time, these algorithms
tend to be black boxes with potentially many enigmatic knobs to turn. A domain
scientist may rightfully ask just which of the many algorithms to use, which
parameters to tune, and what the results actually mean.

The goal of \texttt{cesium} is to simplify the analysis pipeline so that scientists
can spend less time solving technical computing problems and more time answering
scientific questions. \texttt{cesium} provides a library of feature extraction
techniques inspired by analyses from many scientific disciplines, as well as a
surrounding framework for building and analyzing models from the resulting
feature information using \texttt{scikit-learn} (or potentially other machine
learning tools).

By recording the inputs, parameters, and outputs of previous experiments,
\DUroletitlereference{cesium`} allows researchers to answer new questions that arise out of
previous lines of inquiry. Saved \texttt{cesium} workflows can be applied to new
data as it arrives and shared with collaborators or published so that others
may apply the same beginning-to-end analysis for their own data.

For advanced users or users who wish to delve into the source code corresponding
to a workflow produced through the \texttt{cesium} web front end, we are implementing
the ability to produce a Jupyter notebook \DUrole{cite}{perez2007} from a saved
workflow with a single click. While our goal is to have the front end to be as
robust and flexible as possible, ultimately there will always be special cases
where an analysis requires tools which have not been anticipated, or where the
debugging process requires a more detailed look at the intermediate stages of
the analysis. Exporting a workflow to a runnable notebook provides a more
detailed, lower-level look at how the analysis is being performed, and can also
allow the user to reuse certain steps from a given analysis within any other
Python program.

\subsection{\texttt{cesium} library%
  \label{cesium-library}%
}

The first half of the \texttt{cesium} framework is the back-end Python-based library,
aimed at addressing the following uses cases:\setcounter{listcnt0}{0}
\begin{list}{\arabic{listcnt0}.}
{
\usecounter{listcnt0}
\setlength{\rightmargin}{\leftmargin}
}

\item 

A domain scientist who is comfortable with programming but is \textbf{unfamiliar
with time series analysis or machine learning}.
\item 

A scientist who is experienced with time series analysis but is looking for
\textbf{new features} that can better capture patterns within their data.
\item 

A user of the \texttt{cesium} web front end who realizes they require additional
functionality and wishes to add additional stages to their workflow.\end{list}

Our framework primarily implements \textquotedbl{}feature-based methods\textquotedbl{}, wherein the raw
input time series data is used to compute \textquotedbl{}features\textquotedbl{} that compactly capture the
complexity of the signal space within a lower-dimensional feature space.
Standard machine learning approaches (such as random forests \DUrole{cite}{breiman2001}
and support vector machines \DUrole{cite}{suykens1999}) may then be used for supervised
classification or regression.

\texttt{cesium} allows users to select from a large library of features,
including both general time series features and domain-specific features drawn from
various scientific disciplines. Some specific advantages of the \texttt{cesium}
featurization process include:%
\begin{itemize}

\item 

Support for both regularly and irregularly sampled time series.
\item 

Ability to incorporate measurement errors, which can be provided for each data
point of each time series (if applicable).
\item 

Support for multi-channel data, in which case features are computed separately
for each dimension of the input data.
\end{itemize}

\subsubsection{Example features%
  \label{example-features}%
}

Some \texttt{cesium} features are extremely simple and intuitive: summary statistics
such as maximum/minimum values, mean/median values, and standard deviation or median
absolute deviation are a few such examples. Other features involve
measurement errors if they are available: for example, a mean and standard
deviation that is weighted by measurement errors allows noisy data with
large outliers to be modeled more precisely.\begin{figure}[]\noindent\makebox[\columnwidth][c]{\includegraphics[width=\columnwidth]{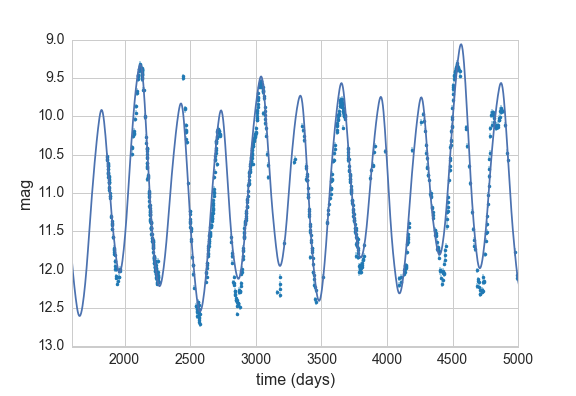}}
\caption{Fitted multi-harmonic Lomb-Scargle model for a light curve from a periodic
Mira-class star. \texttt{cesium} automatically generates numerous features based
on Lomb-Scargle periodogram analysis. \DUrole{label}{ls}}
\end{figure}

Other more involved features could be the estimated parameters for various
fitted statistical models: Figure \DUrole{ref}{ls} shows a multi-frequency,
multi-harmonic Lomb-Scargle model that describes the rich periodic behavior in
an example time series \DUrole{cite}{lomb1976,scargle1982}. The Lomb-Scargle method is
one approach for generalizing the process of Fourier analysis of frequency
spectra to the case of irregularly sampled time series. In particular, a time
series is modeled as a superposition of periodic functions\begin{equation*}
\tilde{y}(t) = \sum_{k=1}^m \sum_{l=1}^n A_{kl} \cos k \omega_l t + B_{kl} \sin k \omega_l t,
\end{equation*}where the parameters $A_{kl}, B_{kl},$ and $\omega_l$ are selected
via non-convex optimization to minimize the residual sum of squares
(weighted by measurement errors if applicable). The estimated periods,
amplitudes, phases, goodness-of-fits, and power spectrum can then be used as
features which broadly characterize the periodicity of the input time series.

\subsubsection{Usage overview%
  \label{usage-overview}%
}

Here we provide a few examples of the main \texttt{cesium} API components that would
be used in a typical analysis task. A workflow will typically consist of three
steps: featurization, model building, and prediction on new data. The majority of
\texttt{cesium} functionality is contained within the \texttt{cesium.featurize} module;
the \texttt{cesium.build\_model} and \texttt{cesium.predict} modules primarily provide
interfaces between sets of feature data, which contain both feature data and a
variety of metadata about the input time series, and machine learning models
from \texttt{scikit-learn} \DUrole{cite}{pedregosa2011}, which require dense, rectangular
input data. Note that, as \texttt{cesium} is under active development, some of the
following details are subject to change.

The featurization step is performed using one of two main functions:%
\begin{itemize}

\item 

\texttt{featurize\_time\_series(times, values, errors, ...)}%
\begin{itemize}

\item 

Takes in data that is already present in memory and computes the requested
features (passed in as string feature names) for each time series.
\item 

Features can be computed in parallel across workers via Celery, a Python
distributed task queue \DUrole{cite}{celery}, or locally in serial.
\item 

Class labels/regression targets and metadata/features with known values are
passed in and stored in the output dataset.
\item 

Additional feature functions can be passed in as \texttt{custom\_functions}.
\end{itemize}

\item 

\texttt{featurize\_data\_files(uris, ...)},%
\begin{itemize}

\item 

Takes in a list of file paths or URIs and dispatches featurization tasks to
be computed in parallel via Celery.
\item 

Data is loaded only remotely by the workers rather than being copied, so
this approach should be preferred for very large input datasets.
\item 

Features, metadata, and custom feature functions are passed in the same way
as \texttt{featurize\_data\_files}.
\end{itemize}

\end{itemize}

The output of both functions is a \texttt{Dataset} object from the \texttt{xarray} library
\DUrole{cite}{xarray}, which will also be referred to here as a \textquotedbl{}feature set\textquotedbl{} (more about
\texttt{xarray} is given in the next section). The feature set stores the computed
feature values for each function (indexed by channel, if the input data is
multi-channel), as well as time series filenames or labels, class labels or
regression targets, and other arbitrary metadata to be used in building a
statistical model.

The \texttt{build\_model} contains tools meant to to simplify the process of building
\texttt{sckit-learn} models from (non-rectangular) feature set data:%
\begin{itemize}

\item 

\texttt{model\_from\_featureset(featureset, ...)}%
\begin{itemize}

\item 

Returns a fitted \texttt{scikit-learn} model based on the input feature data.
\item 

A pre-initialized (but untrained) model can be passed in, or the model type
can be passed in as a string.
\item 

Model parameters can be passed in as fixed values, or as ranges of values
from which to select via cross-validation.
\end{itemize}

\end{itemize}

Analogous helper functions for prediction are available in the \texttt{predict} module:%
\begin{itemize}

\item 

\texttt{model\_predictions(featureset, model, ...)}%
\begin{itemize}

\item 

Generates predictions from a feature set outputted by
\texttt{featurize\_time\_series} or \texttt{featurize\_data\_files}.
\end{itemize}

\item 

\texttt{predict\_data\_files(file\_paths, model, ...)}%
\begin{itemize}

\item 

Like \texttt{featurize\_data\_files}, generate predictions for time series which
have not yet been featurized by dispatching featurization tasks to Celery
workers and then passing the resulting featureset to \texttt{model\_predictions}.
\end{itemize}

\end{itemize}

After a model is initially trained or predictions have been made, new models can
be trained with more features or uninformative features can be removed until the
result is satisfactory.

\subsubsection{Implementation details%
  \label{implementation-details}%
}

\texttt{cesium} is implemented in Python, along with some C code (integrated via
Cython) for especially computationally-intensive feature calculations.
Our library also relies upon many other open source Python projects, including
\texttt{scikit-learn}, \texttt{pandas}, \texttt{xarray}, and \texttt{dask}. As the first two
choices are somewhat obvious, here we briefly describe the roles of the
latter two libraries.

As mentioned above, feature data generated by \texttt{cesium} is returned as a
\texttt{Dataset} object from the \texttt{xarray} package, which according to the
documentation \textquotedbl{}resembles an in-memory representation of a NetCDF file, and
consists of variables, coordinates and attributes which together form a self
describing dataset\textquotedbl{}. A \texttt{Dataset} allows multi-channel feature data to be
faithfully represented in memory as a multidimensional array so that the effects
of each feature (across all channels) or channel (across all features) can be
evaluated directly, while also storing metadata and features that are not
channel-specific. Storing feature outputs in NetCDF format allows for faster and
more space-efficient serialization and loading of results (as compared to a
text-based format).

The \texttt{dask} library provides a wide range of tools for organizing computational
full process of exporting tasks. \texttt{cesium} makes use of only one small
component: within \texttt{dask}, tasks
are organized as a directed acyclic graph (DAG), with the results of some tasks
serving as the inputs to others. Tasks can then be computed in an efficient
order by \texttt{dask}'s scheduler. Within \texttt{cesium}, many features rely on other
features as inputs, so internally we represent our computations as \texttt{dask}
graphs in order to minimize redundant computations and peak memory usage. Part
of an example DAG involving the Lomb-Scargle periodogram is depicted in Figure
\DUrole{ref}{dask}: circles represent functions, and rectangles the inputs/outputs of
the various steps. In addition to the built-in features, custom feature functions
passed in directly by the user can similarly make use of the internal \texttt{dask}
representation so that built-in features can be reused for the evaluation of
user-specified functions.\begin{figure}[]\noindent\makebox[\columnwidth][c]{\includegraphics[scale=0.40]{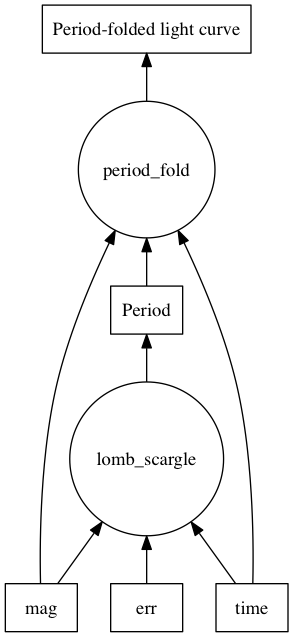}}
\caption{Example of a directed feature computation graph using \texttt{dask}. \DUrole{label}{dask}}
\end{figure}

\subsection{Web front end%
  \label{web-front-end}%
}

The \texttt{cesium} front end provides web-based access to time series
analysis, addressing three common use cases:\setcounter{listcnt0}{0}
\begin{list}{\arabic{listcnt0}.}
{
\usecounter{listcnt0}
\setlength{\rightmargin}{\leftmargin}
}

\item 

A scientist needs to perform time series analysis, but is
\textbf{unfamiliar with programming} and library usage.
\item 

A group of scientists want to \textbf{collaboratively explore} different
methods for time-series analysis.
\item 

A scientist is unfamiliar with time-series analysis, and wants to \textbf{learn}
how to apply various methods to their data, using \textbf{industry best
practices}.\end{list}

The front-end system (together with its deployed back end), offers the
following features:%
\begin{itemize}

\item 

Distributed, parallelized fitting of machine learning models.
\item 

Isolated\DUfootnotemark{id1}{isolation}{1}, cloud-based execution of user-uploaded featurization code.
\item 

Visualization and analysis of results.
\item 

Tracking of an entire exploratory workflow from start-to-finish for
reproducibility (in progress).
\item 

Downloads of Jupyter notebooks to replicate analyses\DUfootnotemark{id2}{notebook}{2}.
\end{itemize}
\DUfootnotetext{isolation}{id1}{1}{\phantomsection\label{isolation}
Isolation is currently provided by limiting the user
to non-privileged access inside a Docker \DUrole{cite}{docker}
container.}
\DUfootnotetext{notebook}{id2}{2}{\phantomsection\label{notebook}
Our current implementation of the front end includes the ability
to track all of a user's actions in order to produce a notebook
version, but the full process of generating the notebook is still
a work in progress.}

\subsubsection{Implementation%
  \label{implementation}%
}

The \texttt{cesium} web front end consists of several components:%
\begin{itemize}

\item 

A Python-based Flask \DUrole{cite}{flask} server which provides a REST API for
managing datasets and launching featurization, model-building, and prediction
tasks.
\item 

A JavaScript-based web interface implemented using React
\DUrole{cite}{gackenheimer2015a} and Redux \DUrole{cite}{gackenheimer2015b} to display results to users.
\item 

A custom WebSocket communication system (which we informally call \emph{message
flow}) that notifies the front end when back-end tasks complete.
\end{itemize}

While the deployment details of the web front end are beyond the scope of this
paper, it should be noted that it was designed with scalability in mind.
The overarching design principle is to connect several small components, each
performing only one, simple task.
An NGINX proxy exposes a pool of WebSocket and Web Server Gateway Interface
(WSGI) servers to the user. This gives us the flexibility to choose the best
implementation of each. Communications between WSGI servers and WebSocket
servers happen through a ZeroMq XPub-XSub (multi-publisher publisher-subscriber)
pipeline \DUrole{cite}{hintjens2013}, but could be replaced with any other broker,
e.g., RabbitMQ \DUrole{cite}{videla2012}. The \textquotedbl{}message flow\textquotedbl{} paradigm adds WebSocket
support to any Python WSGI server (Flask, Django\DUfootnotemark{id3}{channels}{3}, Pylons, etc.), and
allows scaling up as demand increases. It also implement trivially modern data
flow models such as Flux/Redux, where information always flows in one direction:
from front end to back end via HTTP (Hypertext Transfer Protocol) calls, and
from back end to front end via WebSocket communication.%
\DUfootnotetext{channels}{id3}{3}{\phantomsection\label{channels}
At PyCon2016, Andrew Godwin presented a similar
solution for Django called \textquotedbl{}channels\textquotedbl{}. The work
described here happened before we became aware of
Andrew's, and generalizes beyond Django to, e.g.,
Flask, the web framework we use.}

\subsubsection{Computational Scalability%
  \label{computational-scalability}%
}

In many fields, the volumes of available time series data can be immense.
\texttt{cesium} includes features to help parallelize and scale an analysis from a
single system to a large cluster.

Both the back-end library and web front end make use of Celery \DUrole{cite}{celery} for
distributing featurization tasks to multiple workers; this could be used for
anything from automatically utilizing all the available cores of a single machine,
to assigning jobs across a large cluster. Similarly, both parts of the
\texttt{cesium} framework include support for various distributed filesystems, so
that analyses can be performed without copying the entire dataset into a
centralized location.

While the \texttt{cesium} library is written in pure Python, the overhead of the
featurization tasks is minimal; the majority of the work is done by the feature
code itself. Most of the built-in features are based on high-performance
\texttt{numpy} functions; others are written in pure C with interfaces in Cython.
The use of \texttt{dask} graphs to eliminate redundant computations also serves to
minimize memory footprint and reduce computation times.

\subsubsection{Automated testing and documentation%
  \label{automated-testing-and-documentation}%
}

Because the back-end library and web front end are developed in separate GitHub
repositories, the connections between the two somewhat complicate the continuous
integration testing setup. Both repositories are integrated with
\href{https://travis-ci.com/}{Travis CI} for
automatic testing of all branches and pull requests; in addition, any new pushes
to \texttt{cesium/master} trigger a set of tests of the front end using the new
version of the back-end library, with any failures being reported but not
causing the \texttt{cesium} build to fail (the reasoning being that the back-end
library API should be the \textquotedbl{}ground truth\textquotedbl{}, so any updates represent a required
change to the front end, not a bug \emph{per se}).

Documentation for the back-end API is automatically generated in ReStructured
Text format via \texttt{numpydoc}; the result is combined with the rest of our
documentation and rendered as HTML using \texttt{sphinx}. Code examples (without
output) are stored in the repository in Markdown format as opposed to Jupyter
notebooks since this format is better suited to version control. During the
doc-build process, the Markdown is converted to Jupyter notebook format using
\texttt{notedown}, then executed using \texttt{nbconvert} and converted back to Markdown
(with outputs included), to be finally rendered by \texttt{sphinx}.
This allows the code examples to be saved in a human-readable and version
control-friendly format while still allowing the user to execute the code
themselves via a downloadable notebook.

\subsection{Example EEG dataset analysis%
  \label{example-eeg-dataset-analysis}%
}

In this example we compare various techniques for epilepsy detection using a
classic EEG time series dataset from Andrzejak et al. \DUrole{cite}{andrzejak2001}.
The raw data are separated into five classes: Z, O, N, F, and S; we
consider a three-class classification problem of distinguishing normal (Z, O),
interictal (N, F), and ictal (S) signals. We show how to perform the
same analysis using both the back-end Python library and the web front end.
% Here we present an example analysis of a light curve dataset from astronomy
% performed using both the Python library and the equivalent front end workflow.
% The problem involves classifying light curves (i.e., time series consisting
% of times, star brightness values (in magnitudes), and measurement errors) based
% on the type of star from which they were collected. We follow the approach
% of :cite:`` using the same 810 training examples but with a reduced set of features
% for simplicity.

\subsubsection{Python library%
  \label{python-library}%
}

First, we load the data and inspect a representative time series from each class:
Figure \DUrole{ref}{eeg} shows one time series from each of the three classes, after the time
series are loaded from \texttt{cesium.datasets.andrzejak}.

Once the data is loaded, we can generate features for each time series using the
\texttt{cesium.featurize} module. The \texttt{featurize} module includes many built-in choices of
features which can be applied for any type of time series data; here we've chosen a few
generic features that do not have any special biological significance.

If Celery is running, the time series will automatically be split among the available workers
and featurized in parallel; setting \texttt{use\_celery=False} will cause the time series to be
featurized serially.\vspace{1mm}
\begin{Verbatim}[commandchars=\\\{\},fontsize=\footnotesize]
\PY{k+kn}{from} \PY{n+nn}{cesium} \PY{k+kn}{import} \PY{n}{featurize}

\PY{n}{features\PYZus{}to\PYZus{}use} \PY{o}{=} \PY{p}{[}\PY{l+s+s1}{\PYZsq{}}\PY{l+s+s1}{amplitude}\PY{l+s+s1}{\PYZsq{}}\PY{p}{,} \PY{l+s+s1}{\PYZsq{}}\PY{l+s+s1}{maximum}\PY{l+s+s1}{\PYZsq{}}\PY{p}{,}
                   \PY{l+s+s1}{\PYZsq{}}\PY{l+s+s1}{max\PYZus{}slope}\PY{l+s+s1}{\PYZsq{}}\PY{p}{,} \PY{l+s+s1}{\PYZsq{}}\PY{l+s+s1}{median}\PY{l+s+s1}{\PYZsq{}}\PY{p}{,}
                   \PY{l+s+s1}{\PYZsq{}}\PY{l+s+s1}{median\PYZus{}absolute\PYZus{}deviation}\PY{l+s+s1}{\PYZsq{}}\PY{p}{,}
                   \PY{l+s+s1}{\PYZsq{}}\PY{l+s+s1}{percent\PYZus{}beyond\PYZus{}1\PYZus{}std}\PY{l+s+s1}{\PYZsq{}}\PY{p}{,}
                   \PY{l+s+s1}{\PYZsq{}}\PY{l+s+s1}{percent\PYZus{}close\PYZus{}to\PYZus{}median}\PY{l+s+s1}{\PYZsq{}}\PY{p}{,}
                   \PY{l+s+s1}{\PYZsq{}}\PY{l+s+s1}{minimum}\PY{l+s+s1}{\PYZsq{}}\PY{p}{,} \PY{l+s+s1}{\PYZsq{}}\PY{l+s+s1}{skew}\PY{l+s+s1}{\PYZsq{}}\PY{p}{,} \PY{l+s+s1}{\PYZsq{}}\PY{l+s+s1}{std}\PY{l+s+s1}{\PYZsq{}}\PY{p}{,}
                   \PY{l+s+s1}{\PYZsq{}}\PY{l+s+s1}{weighted\PYZus{}average}\PY{l+s+s1}{\PYZsq{}}\PY{p}{]}
\PY{n}{fset\PYZus{}cesium} \PY{o}{=} \PY{n}{featurize}\PY{o}{.}\PY{n}{featurize\PYZus{}time\PYZus{}series}\PY{p}{(}
                  \PY{n}{times}\PY{o}{=}\PY{n}{eeg}\PY{p}{[}\PY{l+s+s2}{\PYZdq{}}\PY{l+s+s2}{times}\PY{l+s+s2}{\PYZdq{}}\PY{p}{]}\PY{p}{,}
                  \PY{n}{values}\PY{o}{=}\PY{n}{eeg}\PY{p}{[}\PY{l+s+s2}{\PYZdq{}}\PY{l+s+s2}{measurements}\PY{l+s+s2}{\PYZdq{}}\PY{p}{]}\PY{p}{,}
                  \PY{n}{errors}\PY{o}{=}\PY{n+nb+bp}{None}\PY{p}{,}
                  \PY{n}{features\PYZus{}to\PYZus{}use}\PY{o}{=}\PY{n}{features\PYZus{}to\PYZus{}use}\PY{p}{,}
                  \PY{n}{targets}\PY{o}{=}\PY{n}{eeg}\PY{p}{[}\PY{l+s+s2}{\PYZdq{}}\PY{l+s+s2}{classes}\PY{l+s+s2}{\PYZdq{}}\PY{p}{]}\PY{p}{)}
\end{Verbatim}
\vspace{1mm}
\vspace{1mm}
\begin{Verbatim}[commandchars=\\\{\},fontsize=\footnotesize]
\PY{o}{\PYZlt{}}\PY{n}{xarray}\PY{o}{.}\PY{n}{Dataset}\PY{o}{\PYZgt{}}
\PY{n}{Dimensions}\PY{p}{:}   \PY{p}{(}\PY{n}{channel}\PY{p}{:} \PY{l+m+mi}{1}\PY{p}{,} \PY{n}{name}\PY{p}{:} \PY{l+m+mi}{500}\PY{p}{)}
\PY{n}{Coordinates}\PY{p}{:}
\PY{o}{*} \PY{n}{channel}   \PY{p}{(}\PY{n}{channel}\PY{p}{)} \PY{n}{int64} \PY{l+m+mi}{0}
\PY{o}{*} \PY{n}{name}      \PY{p}{(}\PY{n}{name}\PY{p}{)} \PY{n}{int64} \PY{l+m+mi}{0} \PY{l+m+mi}{1} \PY{o}{.}\PY{o}{.}\PY{o}{.}
  \PY{n}{target}    \PY{p}{(}\PY{n}{name}\PY{p}{)} \PY{n+nb}{object} \PY{l+s+s1}{\PYZsq{}}\PY{l+s+s1}{Normal}\PY{l+s+s1}{\PYZsq{}} \PY{l+s+s1}{\PYZsq{}}\PY{l+s+s1}{Normal}\PY{l+s+s1}{\PYZsq{}} \PY{o}{.}\PY{o}{.}\PY{o}{.}
\PY{n}{Data} \PY{n}{variables}\PY{p}{:}
  \PY{n}{minimum}   \PY{p}{(}\PY{n}{name}\PY{p}{,} \PY{n}{channel}\PY{p}{)} \PY{n}{float64} \PY{o}{\PYZhy{}}\PY{l+m+mf}{146.0} \PY{o}{\PYZhy{}}\PY{l+m+mf}{254.0} \PY{o}{.}\PY{o}{.}\PY{o}{.}
  \PY{n}{amplitude} \PY{p}{(}\PY{n}{name}\PY{p}{,} \PY{n}{channel}\PY{p}{)} \PY{n}{float64} \PY{l+m+mf}{143.5} \PY{l+m+mf}{211.5} \PY{o}{.}\PY{o}{.}\PY{o}{.}
  \PY{o}{.}\PY{o}{.}\PY{o}{.}
\end{Verbatim}
\vspace{1mm}
The resulting \texttt{Dataset} contains all the feature information needed to train a
machine learning model: feature values are stored as data variables, and the
time series index/class label are stored as coordinates (a \texttt{channel}
coordinate will also be used later for multi-channel data).

Custom feature functions not built into \texttt{cesium} may be passed in using the
\texttt{custom\_functions} keyword, either as a dictionary \texttt{\{feature\_name: function\}}, or as a
\texttt{dask} graph. Functions should take three arrays \texttt{times, measurements, errors} as
inputs; details can be found in the \texttt{cesium.featurize} documentation. Here we
compute five standard features for EEG analysis suggested by Guo et al. \DUrole{cite}{guo2011}:\vspace{1mm}
\begin{Verbatim}[commandchars=\\\{\},fontsize=\footnotesize]
\PY{k+kn}{import} \PY{n+nn}{numpy} \PY{k+kn}{as} \PY{n+nn}{np}\PY{o}{,} \PY{n+nn}{scipy.stats}

\PY{k}{def} \PY{n+nf}{mean\PYZus{}signal}\PY{p}{(}\PY{n}{t}\PY{p}{,} \PY{n}{m}\PY{p}{,} \PY{n}{e}\PY{p}{)}\PY{p}{:}
    \PY{k}{return} \PY{n}{np}\PY{o}{.}\PY{n}{mean}\PY{p}{(}\PY{n}{m}\PY{p}{)}

\PY{k}{def} \PY{n+nf}{std\PYZus{}signal}\PY{p}{(}\PY{n}{t}\PY{p}{,} \PY{n}{m}\PY{p}{,} \PY{n}{e}\PY{p}{)}\PY{p}{:}
    \PY{k}{return} \PY{n}{np}\PY{o}{.}\PY{n}{std}\PY{p}{(}\PY{n}{m}\PY{p}{)}

\PY{k}{def} \PY{n+nf}{mean\PYZus{}square\PYZus{}signal}\PY{p}{(}\PY{n}{t}\PY{p}{,} \PY{n}{m}\PY{p}{,} \PY{n}{e}\PY{p}{)}\PY{p}{:}
    \PY{k}{return} \PY{n}{np}\PY{o}{.}\PY{n}{mean}\PY{p}{(}\PY{n}{m} \PY{o}{*}\PY{o}{*} \PY{l+m+mi}{2}\PY{p}{)}

\PY{k}{def} \PY{n+nf}{abs\PYZus{}diffs\PYZus{}signal}\PY{p}{(}\PY{n}{t}\PY{p}{,} \PY{n}{m}\PY{p}{,} \PY{n}{e}\PY{p}{)}\PY{p}{:}
    \PY{k}{return} \PY{n}{np}\PY{o}{.}\PY{n}{sum}\PY{p}{(}\PY{n}{np}\PY{o}{.}\PY{n}{abs}\PY{p}{(}\PY{n}{np}\PY{o}{.}\PY{n}{diff}\PY{p}{(}\PY{n}{m}\PY{p}{)}\PY{p}{)}\PY{p}{)}

\PY{k}{def} \PY{n+nf}{skew\PYZus{}signal}\PY{p}{(}\PY{n}{t}\PY{p}{,} \PY{n}{m}\PY{p}{,} \PY{n}{e}\PY{p}{)}\PY{p}{:}
    \PY{k}{return} \PY{n}{scipy}\PY{o}{.}\PY{n}{stats}\PY{o}{.}\PY{n}{skew}\PY{p}{(}\PY{n}{m}\PY{p}{)}
\end{Verbatim}
\vspace{1mm}
Now we pass the desired feature functions as a dictionary via the \texttt{custom\_functions}
keyword argument (functions can also be passed in as a list or a \texttt{dask} graph).\vspace{1mm}
\begin{Verbatim}[commandchars=\\\{\},fontsize=\footnotesize]
\PY{n}{guo\PYZus{}features} \PY{o}{=} \PY{p}{\PYZob{}}
    \PY{l+s+s1}{\PYZsq{}}\PY{l+s+s1}{mean}\PY{l+s+s1}{\PYZsq{}}\PY{p}{:} \PY{n}{mean\PYZus{}signal}\PY{p}{,}
    \PY{l+s+s1}{\PYZsq{}}\PY{l+s+s1}{std}\PY{l+s+s1}{\PYZsq{}}\PY{p}{:} \PY{n}{std\PYZus{}signal}\PY{p}{,}
    \PY{l+s+s1}{\PYZsq{}}\PY{l+s+s1}{mean2}\PY{l+s+s1}{\PYZsq{}}\PY{p}{:} \PY{n}{mean\PYZus{}square\PYZus{}signal}\PY{p}{,}
    \PY{l+s+s1}{\PYZsq{}}\PY{l+s+s1}{abs\PYZus{}diffs}\PY{l+s+s1}{\PYZsq{}}\PY{p}{:} \PY{n}{abs\PYZus{}diffs\PYZus{}signal}\PY{p}{,}
    \PY{l+s+s1}{\PYZsq{}}\PY{l+s+s1}{skew}\PY{l+s+s1}{\PYZsq{}}\PY{p}{:} \PY{n}{skew\PYZus{}signal}
\PY{p}{\PYZcb{}}
\PY{n}{fset\PYZus{}guo} \PY{o}{=} \PY{n}{featurize}\PY{o}{.}\PY{n}{featurize\PYZus{}time\PYZus{}series}\PY{p}{(}
               \PY{n}{times}\PY{o}{=}\PY{n}{eeg}\PY{p}{[}\PY{l+s+s2}{\PYZdq{}}\PY{l+s+s2}{times}\PY{l+s+s2}{\PYZdq{}}\PY{p}{]}\PY{p}{,}
               \PY{n}{values}\PY{o}{=}\PY{n}{eeg}\PY{p}{[}\PY{l+s+s2}{\PYZdq{}}\PY{l+s+s2}{measurements}\PY{l+s+s2}{\PYZdq{}}\PY{p}{]}\PY{p}{,}
               \PY{n}{errors}\PY{o}{=}\PY{n+nb+bp}{None}\PY{p}{,} \PY{n}{targets}\PY{o}{=}\PY{n}{eeg}\PY{p}{[}\PY{l+s+s2}{\PYZdq{}}\PY{l+s+s2}{classes}\PY{l+s+s2}{\PYZdq{}}\PY{p}{]}\PY{p}{,}
               \PY{n}{features\PYZus{}to\PYZus{}use}\PY{o}{=}\PY{n}{guo\PYZus{}features}\PY{o}{.}\PY{n}{keys}\PY{p}{(}\PY{p}{)}\PY{p}{,}
               \PY{n}{custom\PYZus{}functions}\PY{o}{=}\PY{n}{guo\PYZus{}features}\PY{p}{)}
\end{Verbatim}
\vspace{1mm}
\vspace{1mm}
\begin{Verbatim}[commandchars=\\\{\},fontsize=\footnotesize]
\PY{o}{\PYZlt{}}\PY{n}{xarray}\PY{o}{.}\PY{n}{Dataset}\PY{o}{\PYZgt{}}
\PY{n}{Dimensions}\PY{p}{:}    \PY{p}{(}\PY{n}{channel}\PY{p}{:} \PY{l+m+mi}{1}\PY{p}{,} \PY{n}{name}\PY{p}{:} \PY{l+m+mi}{500}\PY{p}{)}
\PY{n}{Coordinates}\PY{p}{:}
\PY{o}{*} \PY{n}{channel}    \PY{p}{(}\PY{n}{channel}\PY{p}{)} \PY{n}{int64} \PY{l+m+mi}{0}
\PY{o}{*} \PY{n}{name}       \PY{p}{(}\PY{n}{name}\PY{p}{)} \PY{n}{int64} \PY{l+m+mi}{0} \PY{l+m+mi}{1} \PY{o}{.}\PY{o}{.}\PY{o}{.}
  \PY{n}{target}     \PY{p}{(}\PY{n}{name}\PY{p}{)} \PY{n+nb}{object} \PY{l+s+s1}{\PYZsq{}}\PY{l+s+s1}{Normal}\PY{l+s+s1}{\PYZsq{}} \PY{l+s+s1}{\PYZsq{}}\PY{l+s+s1}{Normal}\PY{l+s+s1}{\PYZsq{}} \PY{o}{.}\PY{o}{.}\PY{o}{.}
\PY{n}{Data} \PY{n}{variables}\PY{p}{:}
  \PY{n}{abs\PYZus{}diffs}  \PY{p}{(}\PY{n}{name}\PY{p}{,} \PY{n}{channel}\PY{p}{)} \PY{n}{float64} \PY{l+m+mf}{4695.2} \PY{l+m+mf}{6112.6} \PY{o}{.}\PY{o}{.}\PY{o}{.}
  \PY{n}{mean}       \PY{p}{(}\PY{n}{name}\PY{p}{,} \PY{n}{channel}\PY{p}{)} \PY{n}{float64} \PY{o}{\PYZhy{}}\PY{l+m+mf}{4.132} \PY{o}{\PYZhy{}}\PY{l+m+mf}{52.44} \PY{o}{.}\PY{o}{.}\PY{o}{.}
  \PY{o}{.}\PY{o}{.}\PY{o}{.}
\end{Verbatim}
\vspace{1mm}
The EEG time series considered here consist of univariate signal measurements along a
uniform time grid. But \texttt{featurize\_time\_series} also accepts multi-channel data. To
demonstrate this, we will decompose each signal into five frequency bands using a discrete
wavelet transform as suggested by Subasi \DUrole{cite}{subasi2007}, and then featurize each band
separately using the five functions from above.\vspace{1mm}
\begin{Verbatim}[commandchars=\\\{\},fontsize=\footnotesize]
\PY{k+kn}{import} \PY{n+nn}{pywt}

\PY{n}{eeg}\PY{p}{[}\PY{l+s+s2}{\PYZdq{}}\PY{l+s+s2}{dwts}\PY{l+s+s2}{\PYZdq{}}\PY{p}{]} \PY{o}{=} \PY{p}{[}\PY{n}{pywt}\PY{o}{.}\PY{n}{wavedec}\PY{p}{(}\PY{n}{m}\PY{p}{,} \PY{n}{pywt}\PY{o}{.}\PY{n}{Wavelet}\PY{p}{(}\PY{l+s+s1}{\PYZsq{}}\PY{l+s+s1}{db1}\PY{l+s+s1}{\PYZsq{}}\PY{p}{)}\PY{p}{,}
                            \PY{n}{level}\PY{o}{=}\PY{l+m+mi}{4}\PY{p}{)}
               \PY{k}{for} \PY{n}{m} \PY{o+ow}{in} \PY{n}{eeg}\PY{p}{[}\PY{l+s+s2}{\PYZdq{}}\PY{l+s+s2}{measurements}\PY{l+s+s2}{\PYZdq{}}\PY{p}{]}\PY{p}{]}
\PY{n}{fset\PYZus{}dwt} \PY{o}{=} \PY{n}{featurize}\PY{o}{.}\PY{n}{featurize\PYZus{}time\PYZus{}series}\PY{p}{(}
               \PY{n}{times}\PY{o}{=}\PY{n+nb+bp}{None}\PY{p}{,} \PY{n}{values}\PY{o}{=}\PY{n}{eeg}\PY{p}{[}\PY{l+s+s2}{\PYZdq{}}\PY{l+s+s2}{dwts}\PY{l+s+s2}{\PYZdq{}}\PY{p}{]}\PY{p}{,} \PY{n}{errors}\PY{o}{=}\PY{n+nb+bp}{None}\PY{p}{,}
               \PY{n}{features\PYZus{}to\PYZus{}use}\PY{o}{=}\PY{n}{guo\PYZus{}features}\PY{o}{.}\PY{n}{keys}\PY{p}{(}\PY{p}{)}\PY{p}{,}
               \PY{n}{targets}\PY{o}{=}\PY{n}{eeg}\PY{p}{[}\PY{l+s+s2}{\PYZdq{}}\PY{l+s+s2}{classes}\PY{l+s+s2}{\PYZdq{}}\PY{p}{]}\PY{p}{,}
               \PY{n}{custom\PYZus{}functions}\PY{o}{=}\PY{n}{guo\PYZus{}features}\PY{p}{)}
\end{Verbatim}
\vspace{1mm}
\vspace{1mm}
\begin{Verbatim}[commandchars=\\\{\},fontsize=\footnotesize]
\PY{o}{\PYZlt{}}\PY{n}{xarray}\PY{o}{.}\PY{n}{Dataset}\PY{o}{\PYZgt{}}
\PY{n}{Dimensions}\PY{p}{:}    \PY{p}{(}\PY{n}{channel}\PY{p}{:} \PY{l+m+mi}{5}\PY{p}{,} \PY{n}{name}\PY{p}{:} \PY{l+m+mi}{500}\PY{p}{)}
\PY{n}{Coordinates}\PY{p}{:}
\PY{o}{*} \PY{n}{channel}    \PY{p}{(}\PY{n}{channel}\PY{p}{)} \PY{n}{int64} \PY{l+m+mi}{0} \PY{l+m+mi}{1} \PY{o}{.}\PY{o}{.}\PY{o}{.}
\PY{o}{*} \PY{n}{name}       \PY{p}{(}\PY{n}{name}\PY{p}{)} \PY{n}{int64} \PY{l+m+mi}{0} \PY{l+m+mi}{1} \PY{o}{.}\PY{o}{.}\PY{o}{.}
  \PY{n}{target}     \PY{p}{(}\PY{n}{name}\PY{p}{)} \PY{n+nb}{object} \PY{l+s+s1}{\PYZsq{}}\PY{l+s+s1}{Normal}\PY{l+s+s1}{\PYZsq{}} \PY{l+s+s1}{\PYZsq{}}\PY{l+s+s1}{Normal}\PY{l+s+s1}{\PYZsq{}} \PY{o}{.}\PY{o}{.}\PY{o}{.}
\PY{n}{Data} \PY{n}{variables}\PY{p}{:}
  \PY{n}{abs\PYZus{}diffs}  \PY{p}{(}\PY{n}{name}\PY{p}{,} \PY{n}{channel}\PY{p}{)} \PY{n}{float64} \PY{l+m+mi}{25131} \PY{l+m+mi}{18069} \PY{o}{.}\PY{o}{.}\PY{o}{.}
  \PY{n}{skew}       \PY{p}{(}\PY{n}{name}\PY{p}{,} \PY{n}{channel}\PY{p}{)} \PY{n}{float64} \PY{o}{\PYZhy{}}\PY{l+m+mf}{0.0433} \PY{l+m+mf}{0.06578} \PY{o}{.}\PY{o}{.}\PY{o}{.}
  \PY{o}{.}\PY{o}{.}\PY{o}{.}
\end{Verbatim}
\vspace{1mm}
The output feature set has the same form as before, except now the \texttt{channel} coordinate is
used to index the features by the corresponding frequency band. The functions in
\texttt{cesium.build\_model} and \texttt{cesium.predict} all accept feature sets from
single- or multi-channel data, so no additional steps are required to train
models or make predictions for multichannel feature sets using the \texttt{cesium}
library.

Model building in \texttt{cesium} is handled by the \texttt{model\_from\_featureset}
function in the \texttt{cesium.build\_model} module. The feature set output by
\texttt{featurize\_time\_series} contains both the feature and target information
needed to train a model; \texttt{model\_from\_featureset} is simply a wrapper
that calls the \texttt{fit} method of a given \texttt{scikit-learn} model with the
appropriate inputs. In the case of multichannel features, it also handles
reshaping the feature set into a (rectangular) form that is compatible with
\texttt{scikit-learn}.

For this example, we test a random forest classifier for the built-in \texttt{cesium} features,
and a 3-nearest neighbors classifier for the others, as in \DUrole{cite}{guo2011}.\vspace{1mm}
\begin{Verbatim}[commandchars=\\\{\},fontsize=\footnotesize]
\PY{k+kn}{from} \PY{n+nn}{cesium.build\PYZus{}model} \PY{k+kn}{import} \PY{n}{model\PYZus{}from\PYZus{}featureset}
\PY{k+kn}{from} \PY{n+nn}{sklearn.ensemble} \PY{k+kn}{import} \PY{n}{RandomForestClassifier}
\PY{k+kn}{from} \PY{n+nn}{sklearn.neighbors} \PY{k+kn}{import} \PY{n}{KNeighborsClassifier}
\PY{k+kn}{from} \PY{n+nn}{sklearn.cross\PYZus{}validation} \PY{k+kn}{import} \PY{n}{train\PYZus{}test\PYZus{}split}

\PY{n}{train}\PY{p}{,} \PY{n}{test} \PY{o}{=} \PY{n}{train\PYZus{}test\PYZus{}split}\PY{p}{(}\PY{l+m+mi}{500}\PY{p}{)}

\PY{n}{rfc\PYZus{}param\PYZus{}grid} \PY{o}{=} \PY{p}{\PYZob{}}\PY{l+s+s1}{\PYZsq{}}\PY{l+s+s1}{n\PYZus{}estimators}\PY{l+s+s1}{\PYZsq{}}\PY{p}{:} \PY{p}{[}\PY{l+m+mi}{8}\PY{p}{,} \PY{l+m+mi}{32}\PY{p}{,} \PY{l+m+mi}{128}\PY{p}{,} \PY{l+m+mi}{512}\PY{p}{]}\PY{p}{\PYZcb{}}
\PY{n}{model\PYZus{}cesium} \PY{o}{=} \PY{n}{model\PYZus{}from\PYZus{}featureset}\PY{p}{(}
                   \PY{n}{fset\PYZus{}cesium}\PY{o}{.}\PY{n}{isel}\PY{p}{(}\PY{n}{name}\PY{o}{=}\PY{n}{train}\PY{p}{)}\PY{p}{,}
                   \PY{n}{RandomForestClassifier}\PY{p}{(}\PY{p}{)}\PY{p}{,}
                   \PY{n}{params\PYZus{}to\PYZus{}optimize}\PY{o}{=}\PY{n}{rfc\PYZus{}param\PYZus{}grid}\PY{p}{)}

\PY{n}{knn\PYZus{}param\PYZus{}grid} \PY{o}{=} \PY{p}{\PYZob{}}\PY{l+s+s1}{\PYZsq{}}\PY{l+s+s1}{n\PYZus{}neighbors}\PY{l+s+s1}{\PYZsq{}}\PY{p}{:} \PY{p}{[}\PY{l+m+mi}{1}\PY{p}{,} \PY{l+m+mi}{2}\PY{p}{,} \PY{l+m+mi}{3}\PY{p}{,} \PY{l+m+mi}{4}\PY{p}{]}\PY{p}{\PYZcb{}}
\PY{n}{model\PYZus{}guo} \PY{o}{=} \PY{n}{model\PYZus{}from\PYZus{}featureset}\PY{p}{(}
                \PY{n}{fset\PYZus{}guo}\PY{o}{.}\PY{n}{isel}\PY{p}{(}\PY{n}{name}\PY{o}{=}\PY{n}{train}\PY{p}{)}\PY{p}{,}
                \PY{n}{KNeighborsClassifier}\PY{p}{(}\PY{p}{)}\PY{p}{,}
                \PY{n}{params\PYZus{}to\PYZus{}optimize}\PY{o}{=}\PY{n}{knn\PYZus{}param\PYZus{}grid}\PY{p}{)}
\PY{n}{model\PYZus{}dwt} \PY{o}{=} \PY{n}{model\PYZus{}from\PYZus{}featureset}\PY{p}{(}
                \PY{n}{fset\PYZus{}dwt}\PY{o}{.}\PY{n}{isel}\PY{p}{(}\PY{n}{name}\PY{o}{=}\PY{n}{train}\PY{p}{)}\PY{p}{,}
                \PY{n}{KNeighborsClassifier}\PY{p}{(}\PY{p}{)}\PY{p}{,}
                \PY{n}{params\PYZus{}to\PYZus{}optimize}\PY{o}{=}\PY{n}{knn\PYZus{}param\PYZus{}grid}\PY{p}{)}
\end{Verbatim}
\vspace{1mm}
Making predictions for new time series based on these models follows the same pattern:
first the time series are featurized using
\texttt{featurize\_timeseries}
and then predictions are made based on these features using
\texttt{predict.model\_predictions},\vspace{1mm}
\begin{Verbatim}[commandchars=\\\{\},fontsize=\footnotesize]
\PY{k+kn}{from} \PY{n+nn}{cesium.predict} \PY{k+kn}{import} \PY{n}{model\PYZus{}predictions}
\PY{n}{preds\PYZus{}cesium} \PY{o}{=} \PY{n}{model\PYZus{}predictions}\PY{p}{(}
                   \PY{n}{fset\PYZus{}cesium}\PY{p}{,} \PY{n}{model\PYZus{}cesium}\PY{p}{,}
                   \PY{n}{return\PYZus{}probs}\PY{o}{=}\PY{n+nb+bp}{False}\PY{p}{)}
\PY{n}{preds\PYZus{}guo} \PY{o}{=} \PY{n}{model\PYZus{}predictions}\PY{p}{(}\PY{n}{fset\PYZus{}guo}\PY{p}{,} \PY{n}{model\PYZus{}guo}\PY{p}{,}
                   \PY{n}{return\PYZus{}probs}\PY{o}{=}\PY{n+nb+bp}{False}\PY{p}{)}
\PY{n}{preds\PYZus{}dwt} \PY{o}{=} \PY{n}{model\PYZus{}predictions}\PY{p}{(}\PY{n}{fset\PYZus{}dwt}\PY{p}{,} \PY{n}{model\PYZus{}dwt}\PY{p}{,}
                   \PY{n}{return\PYZus{}probs}\PY{o}{=}\PY{n+nb+bp}{False}\PY{p}{)}
\end{Verbatim}
\vspace{1mm}
And finally, checking the accuracy of our various models, we find:\vspace{1mm}
\begin{Verbatim}[commandchars=\\\{\},fontsize=\footnotesize]
\PY{n}{Builtin}\PY{p}{:} \PY{n}{train} \PY{n}{acc}\PY{o}{=}\PY{l+m+mf}{100.00}\PY{o}{\PYZpc{}}\PY{p}{,} \PY{n}{test} \PY{n}{acc}\PY{o}{=}\PY{l+m+mf}{83.20}\PY{o}{\PYZpc{}}
\PY{n}{Guo} \PY{n}{et} \PY{n}{al}\PY{o}{.}\PY{p}{:} \PY{n}{train} \PY{n}{acc}\PY{o}{=}\PY{l+m+mf}{90.93}\PY{o}{\PYZpc{}}\PY{p}{,} \PY{n}{test} \PY{n}{acc}\PY{o}{=}\PY{l+m+mf}{84.80}\PY{o}{\PYZpc{}}
\PY{n}{Wavelets}\PY{p}{:} \PY{n}{train} \PY{n}{acc}\PY{o}{=}\PY{l+m+mf}{100.00}\PY{o}{\PYZpc{}}\PY{p}{,} \PY{n}{test} \PY{n}{acc}\PY{o}{=}\PY{l+m+mf}{95.20}\PY{o}{\PYZpc{}}
\end{Verbatim}
\vspace{1mm}
The workflow presented here is intentionally simplistic and omits many important steps
such as feature selection, model parameter selection, etc., which may all be
incorporated just as they would for any other \texttt{scikit-learn} analysis.
But with essentially three function calls (\texttt{featurize\_time\_series},
\texttt{model\_from\_featureset}, and \texttt{model\_predictions}), we are able to build a
model from a set of time series and make predictions on new, unlabeled data. In
the next section we introduce the web front end for \texttt{cesium} and describe how
the same analysis can be performed in a browser with no setup or coding required.

\subsubsection{Web front end%
  \label{id4}%
}

Here we briefly demonstrate how the above analysis could be conducted using only
the web front end. Note that the user interface presented here is a preliminary version
and is undergoing frequent updates and additions. The basic workflow follows the
same \emph{featurize}—\emph{build model}—\emph{predict} pattern. First,
data is uploaded as in Figure \DUrole{ref}{web2}. Features are
selected from available built-in functions as in Figure \DUrole{ref}{web3},
or may be computed from user-uploaded Python code which is securely executed
within a Docker container. Once features have been extracted, models can be
created as in Figure \DUrole{ref}{web4}, and finally predictions can be made as in
Figure \DUrole{ref}{web5}. Currently the options for exploring feature importance and
model accuracy are limited, but this is again an area of active development.\begin{figure}[]\noindent\makebox[\columnwidth][c]{\includegraphics[width=\columnwidth]{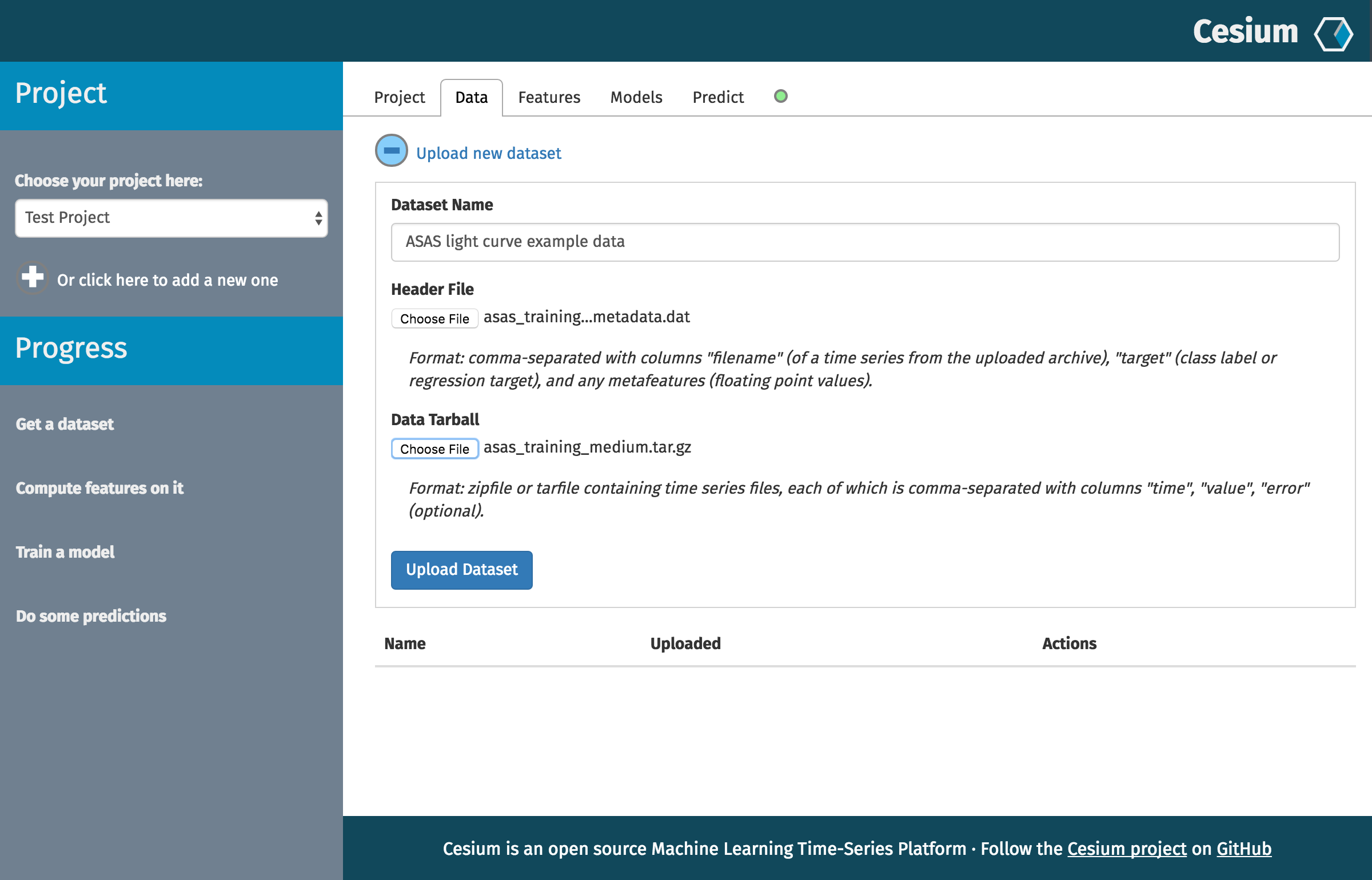}}
\caption{\textquotedbl{}Data\textquotedbl{} tab \DUrole{label}{web2}}
\end{figure}\begin{figure}[]\noindent\makebox[\columnwidth][c]{\includegraphics[width=\columnwidth]{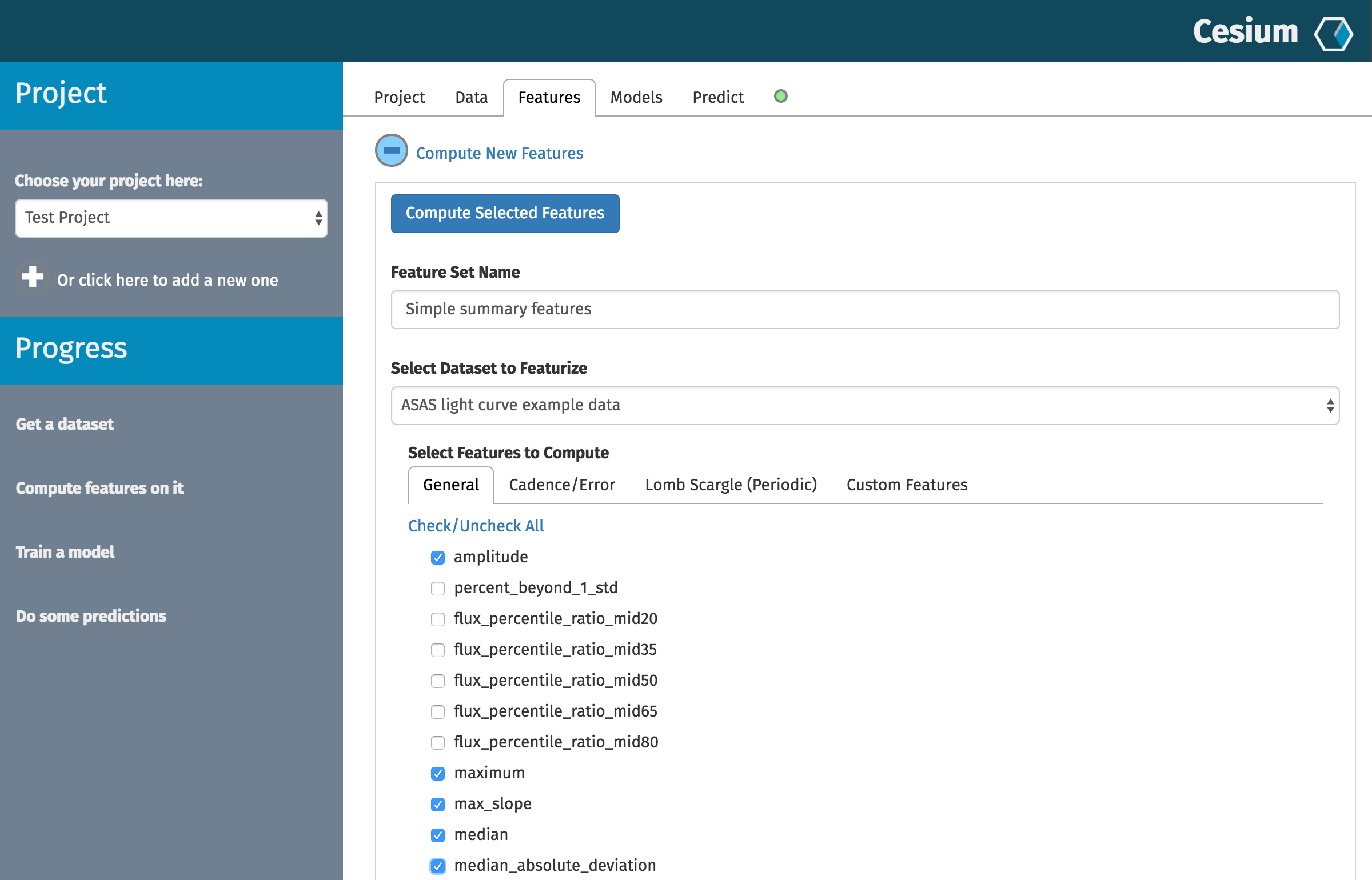}}
\caption{\textquotedbl{}Featurize\textquotedbl{} tab \DUrole{label}{web3}}
\end{figure}\begin{figure}[]\noindent\makebox[\columnwidth][c]{\includegraphics[width=\columnwidth]{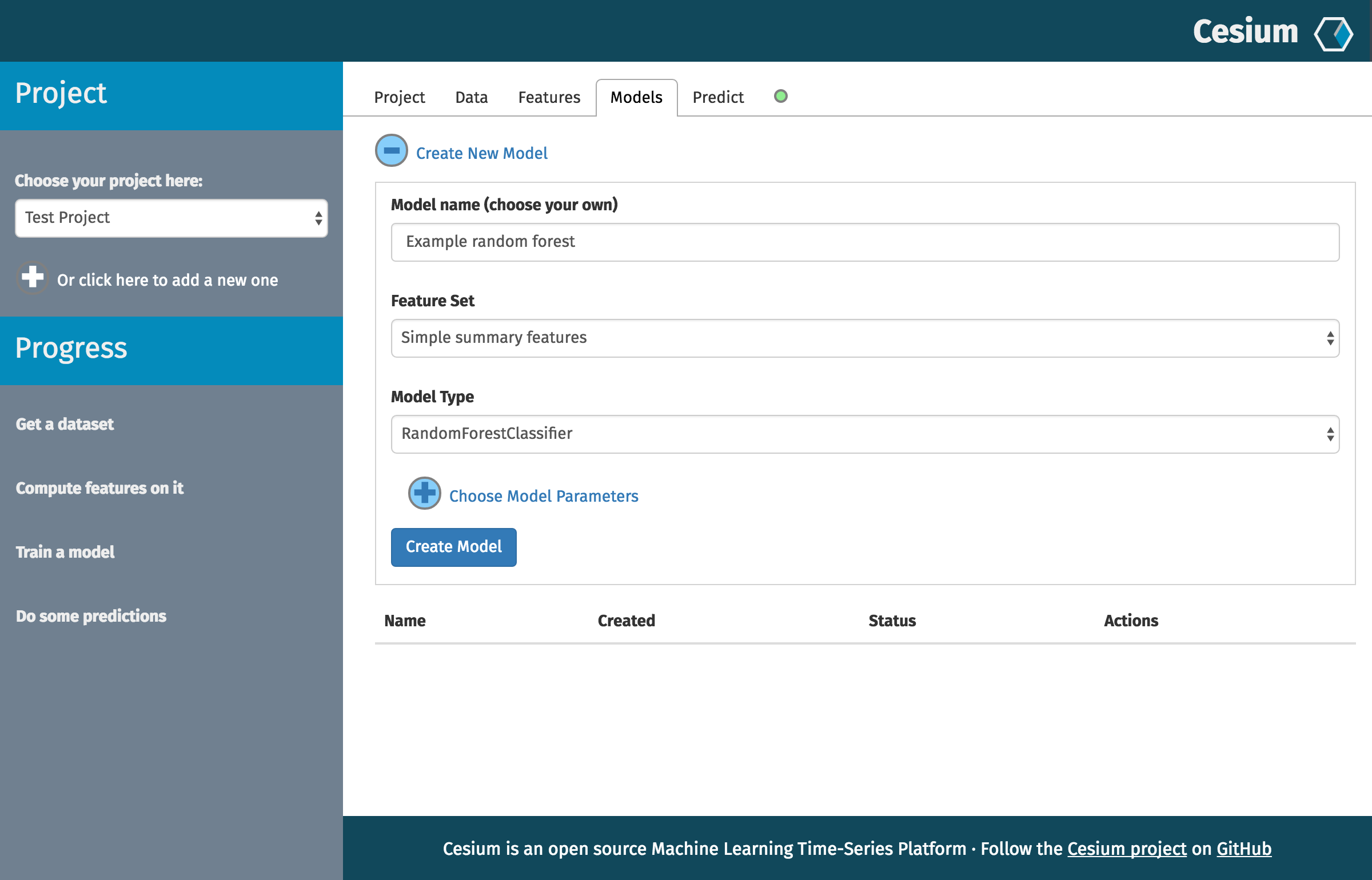}}
\caption{\textquotedbl{}Build Model\textquotedbl{} tab \DUrole{label}{web4}}
\end{figure}\begin{figure}[]\noindent\makebox[\columnwidth][c]{\includegraphics[width=\columnwidth]{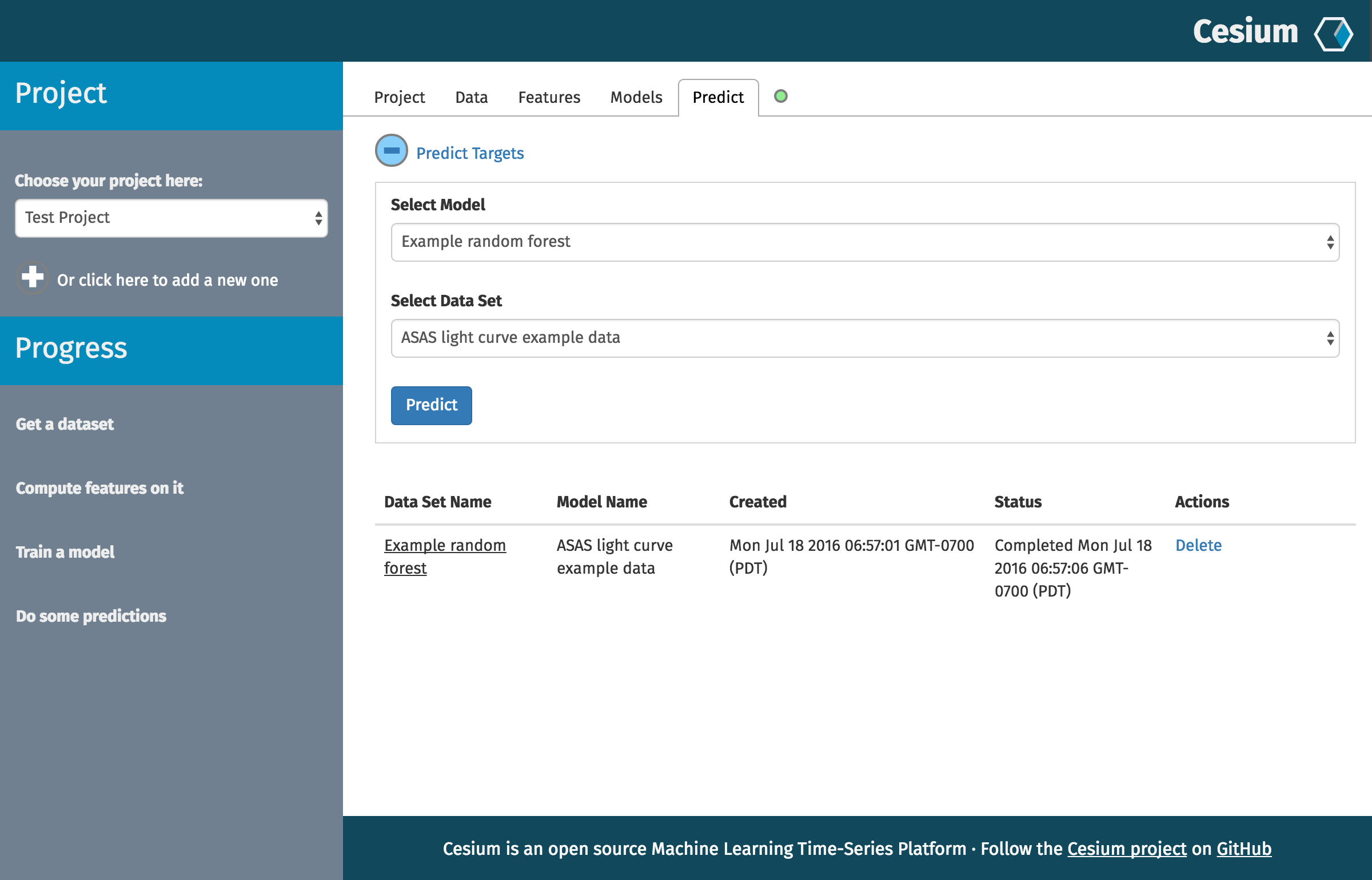}}
\caption{\textquotedbl{}Predict\textquotedbl{} tab \DUrole{label}{web5}}
\end{figure}

\subsection{Future work%
  \label{future-work}%
}

The \texttt{cesium} project is under active development. Some of our upcoming goals
include:%
\begin{itemize}

\item 

Full support for exporting Jupyter notebooks from any workflow created within
the web front end.
\item 

Additional features from other scientific disciplines (currently the majority
of available features are taken from applications in astronomy).
\item 

Improved web front end user interface with more tools for visualizing and
exploring a user's raw data, feature values, and model outputs.
\item 

More tools to streamline the process of iteratively exploring new models based
on results of previous experiments.
\item 

Better support for sharing data and results among teams.
\item 

Extension to unsupervised problems.
\end{itemize}

\subsection{Conclusion%
  \label{conclusion}%
}

The \texttt{cesium} framework provides tools that allow anyone from machine learning
specialists to domain experts without any machine learning experience to rapidly
prototype explanatory models for their time series data and generate predictions
for new, unlabeled data. Aside from the applications to time domain informatics,
our project has several aspects which are relevant to the broader scientific
Python community.

First, the dual nature of the project (Python back end vs. web front end) presents
both unique challenges and interesting opportunities in striking a balance
between accessibility and flexibility of the two components.
Second, the \texttt{cesium} project places a strong emphasis on reproducible
workflows: all actions performed within the web front end are logged and can be
easily exported to a Jupyter notebook that exactly reproduces the steps of the
analysis. Finally, the scope of our project is simultaneously both narrow (time
series analysis) and broad (numerous distinct scientific disciplines), so
determining how much domain-specific functionality to include is an ongoing
challenge.

\subsection{Acknowledgements%
  \label{acknowledgements}%
}

The authors acknowledge the generous support of BIGDATA grant \#1251274 from the
National Science Foundation and the Data-Driven Investigator Award from the
Gordon and Betty Moore Foundation.
\bibliographystyle{alphaurl}
\bibliography{mybib}

\end{document}